\documentclass[apj]{emulateapj}

\usepackage{color}

\shorttitle{Dust-trapping vortices and a potentially planet-triggered spiral wake in the disk of V1247\,Orionis}
\shortauthors{Kraus et al.}

\begin{document}


\title{Dust-trapping vortices and a potentially planet-triggered spiral wake in the pre-transitional disk of V1247\,Orionis}


\author{
Stefan Kraus\altaffilmark{1}, 
Alexander Kreplin\altaffilmark{1},
Misato Fukugawa\altaffilmark{2},
Takayuki Muto\altaffilmark{3}, 
Michael L.\ Sitko\altaffilmark{4},
Alison K.\ Young\altaffilmark{1},
Matthew R.\ Bate\altaffilmark{1}, 
Carol Grady\altaffilmark{5}, 
Tim T.\ Harries\altaffilmark{1},
John D.\ Monnier\altaffilmark{6},
Matthew Willson\altaffilmark{1}, 
John Wisniewski\altaffilmark{7}
}

\email{skraus@astro.ex.ac.uk}

\affil{
$^{1}$~University of Exeter, School of Physics, Astrophysics Group, Stocker Road, Exeter, EX4 4QL, UK\\
$^{2}$~Division of Particle and Astrophysical Science, Graduate School of Science, Nagoya University, Nagoya, Japan\\
$^{3}$~Division of Liberal Arts, Kogakuin University, 1-24-2 Nishi-Shinjuku, Shinjuku-ku, Tokyo, 163-8677, Japan\\
$^{4}$~Department of Physics, University of Cincinnati, Cincinnati, OH 45221, USA; Center for Extrasolar Planetary Systems, Space Science Institute, 4750 Walnut Street, Suite 205, Boulder, CO 80301, USA\\
$^{5}$~Eureka Scientific, 2452 Delmer Street, Suite 100, Oakland CA 96402, USA\\
$^{6}$~Department of Astronomy, University of Michigan, 311 West Hall, 1085 South University Ave, Ann Arbor, MI 48109, USA\\
$^{7}$~Homer L.\ Dodge Department of Physics, University of Oklahoma, Norman, OK 73071, USA
}

\begin{abstract}
The radial drift problem constitutes one of the most fundamental problems in planet formation theory,
as it predicts particles to drift into the star before they are able to grow to planetesimal size.
Dust-trapping vortices have been proposed as a possible solution to this problem,
as they might be able to trap particles over millions of years, allowing them to grow beyond the radial drift barrier.
Here, we present ALMA {0.04\arcsec}-resolution imaging of the pre-transitional disk of 
V1247\,Orionis that reveals an asymmetric ring as well as a sharply-confined crescent structure,
resembling morphologies seen in theoretical models of vortex formation.
The asymmetric ring (at {0.17\arcsec}=54\,au separation from the star) and the crescent (at {0.38\arcsec}=120\,au) 
seem smoothly connected through a one-armed spiral arm structure that has been found previously in scattered light.
We propose a physical scenario with a planet orbiting at {$\sim0.3$\arcsec$\approx$100\,au},
where the one-armed spiral arm detected in polarised light traces the accretion stream 
feeding the protoplanet. 
The dynamical influence of the planet clears the gap between the ring and the crescent 
and triggers two vortices that trap mm-sized particles, namely the crescent and the bright asymmetry seen in the ring.
We conducted dedicated hydrodynamics simulations of a disk with an embedded planet, 
which results in similar spiral-arm morphologies as seen in our scattered light images. 
{At the position of the spiral wake and the crescent we also observe $^{12}$CO\,(3-2) and 
H$^{12}$CO$^{+}$\,(4-3) excess line emission, likely tracing the increased scale-height in these disk regions.}
\end{abstract}



\section{Introduction}

One of the major unsolved problems in the field of planet formation is the radial drift problem, 
where drag forces between the dust and gas cause the orbits of dust grains to decay \citep{wei77}.
The radial drift velocity scales with the particle size and might prevent dust grains from growing
beyond millimeter size (at 100\,au separation).
Larger particles would drift into the star before they are able to grow further into planetesimals and planets.
One promising mechanism that has been proposed for solving the radial drift barrier is dust trapping, 
where the dust particles get trapped in a pressure bump and converge azimuthally towards the 
pressure maximum instead of acquiring high inward drift {\citep[e.g.][]{bra08,bir13}}.

\citet{pin12} estimated that a local enhancement of 30\% in gas pressure 
would be sufficient to trap particles effectively over several million years.
Over the last years, several mechanisms have been proposed that could 
produce the required long-lifed vortices in the disk, including
the  presence of planets \citep[e.g.][]{dod11,zhu12,pin12} 
and the strong viscosity-gradients that are encountered at the edges of disk dead zones \citep{reg12}.

Observationally, dust-trapping vortices could be detected at millimeter wavelengths 
as asymmetric structures with a wide azimuthal but narrow radial extend. 
{SMA+CARMA} interferometry already identified several disks with asymmetric 
structures that have been interpreted as dust traps 
\citep{reg12}.
ALMA observations on \object{Oph-IRS48} \citep{van13} 
with {0.22\arcsec} resolution showed a peanut-shaped structure that 
spans less than a third of the ring azimuth, while {0.16\arcsec}-resolution imaging 
with superuniform weighting suggested the presence of a ring+arc structure around 
\object{HD135344B} \citep{van16}.

\object{V1247\,Orionis} is an F0V-type star 
at a distance of $320\pm30$\,pc\footnote{All physical scales from the literature have been 
adjusted to match the GAIA distance adopted in this paper.}
\citep[GAIA data release~1,][]{gaia1a,gaia1b}.
Its SED exhibits a strong near-/far-infrared excess, but only modest mid-infrared excess,
characteristic for pre-transitional disks.  
\citet{kra13} resolved at near-infrared wavelengths (NIR, $1.65-2.5\,\mu$m) a narrow, 
optically-thick inner disk at 0.15-0.22\,au. 
This inner disk is separated from the outer disk by a gap that was resolved
at mid-infrared wavelengths (MIR, $8-12\,\mu$m) with Gemini speckle interferometry
($\gtrsim$38\,au={0.120\arcsec}). 
Modeling the NIR+MIR visibilities and SED revealed the presence of 
optically thin, sub-$\mu$m-sized, carbon-rich dust grains located inside the gap, which might indicate that V1247\,Ori 
is in a particularly early stage of disk clearing \citep{kra13}.
\citet{wil17} analysed Keck/NIRC2+VLT/NACO aperture masking observations
from three epochs and
detected a strong asymmetric structure that moves in position angle (PA) from 
$-52\pm3^{\circ}$ (2012.03), $-7\pm3^{\circ}$ (2012.97), to $38\pm3^{\circ}$ (2013.81), 
consistent with Keplerian motion of a companion on an $\sim5$\,au orbit. 

Finally, V1247\,Ori was imaged with Subaru/HiCIAO polarimetry \citep{oht16}, which revealed
a one-sided spiral arm extending from the inner working angle 
({0.14\arcsec}) to {$\sim0.3$\arcsec} in South-East direction.
The spiral arm features an exceptionally large azimuthal contrast: the radial peak of the south-eastern arc 
is about $3\times$ brighter than the north-western disk measured at the same distance from the star.

Here, we present ALMA observations of \object{V1247\,Orionis} 
that reveal a ring-shaped inner disk component with a prominent asymmetry and a sharply-confined crescent structure.
The observation provide a $5-6\times$ improvement in angular resolution compared to earlier observations
of candidates for dust trapping-vortices. V1247\,Orionis also represents the first case where two vortices might
have been observed in the same system, near the inner/outer edge of a density gap.

\section{Observations}
\label{sec:observations}

\object{V1247\,Ori} was observed on 2015-11-16 (120\,min),
2015-12-01 (68\,min) and 2015-12-05 (134\,min) with 47 antennas of ALMA's 12\,m array.
The array was configured for the longest baselines offered in Cycle~3 (C36-8/7),
with baselines between 82\,m and 11.05\,km.
We observed in ALMA band~7 at frequencies around 320\,GHz (870\,$\mu$m)
with an aggregate bandwidth of 7.27\,GHz.
The high-resolution channels covered the $^{12}$CO\,(3-2) and H$^{12}$CO$^{+}$\,(4-3) line
with a velocity resolution of $\sim0.83$\,km\,s$^{-1}$. 
The sources \object{J0510+1800} and \object{J0522-3627} were used as bandpass
calibrators, \object{J0541-0211} as phase calibrator, and \object{J0541-0541} as flux calibrator.

For our continuum imaging, we excluded all channels that include line emission.  
The image reconstructions were carried out using CASA version 4.5.1-REL and phase self-calibration.
{We computed a grid with different reconstruction parameters and weighting schemes
and selected the image with the lowest root-mean-square noise, using
briggs weighting, robust parameter 0.6, threshold 0.06\,mJy, and a {0.45\arcsec} mask.}

\section{Results}
\label{sec:results}

\subsection{Continuum imaging}
\label{sec:resultscont}

\begin{figure*}
  \centering
  $\begin{array}{cc}
     \multicolumn{2}{c}{\includegraphics[angle=0,scale=0.72]{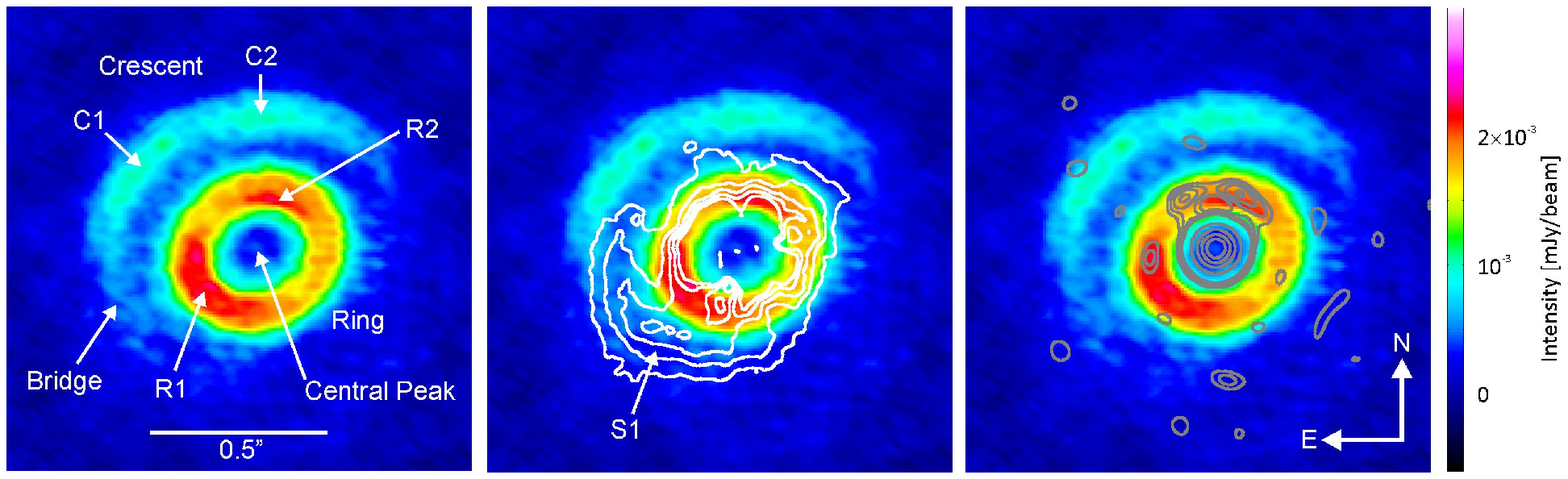}}\\
     \includegraphics[angle=0,scale=0.25]{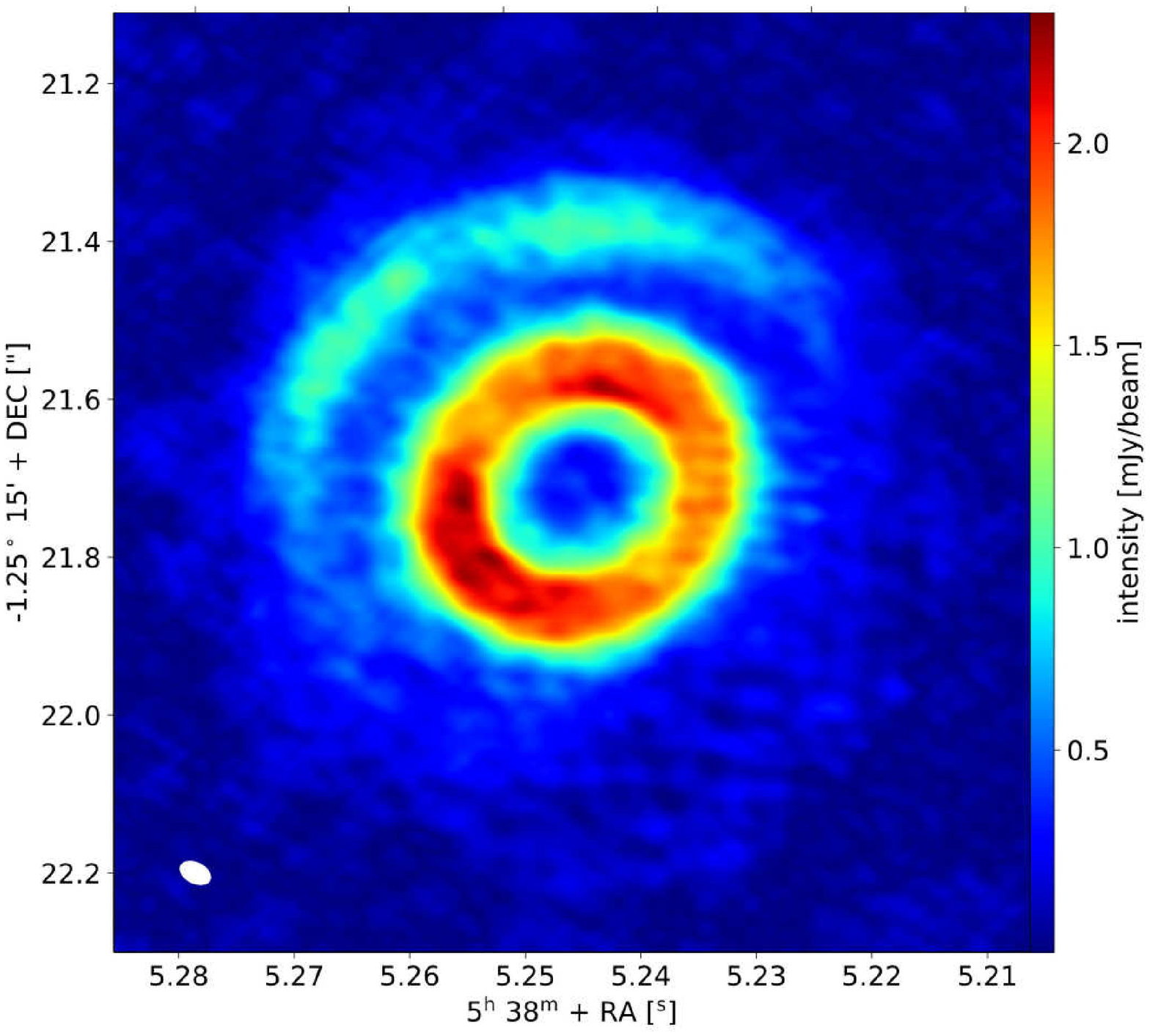} &
     \includegraphics[angle=0,scale=0.25]{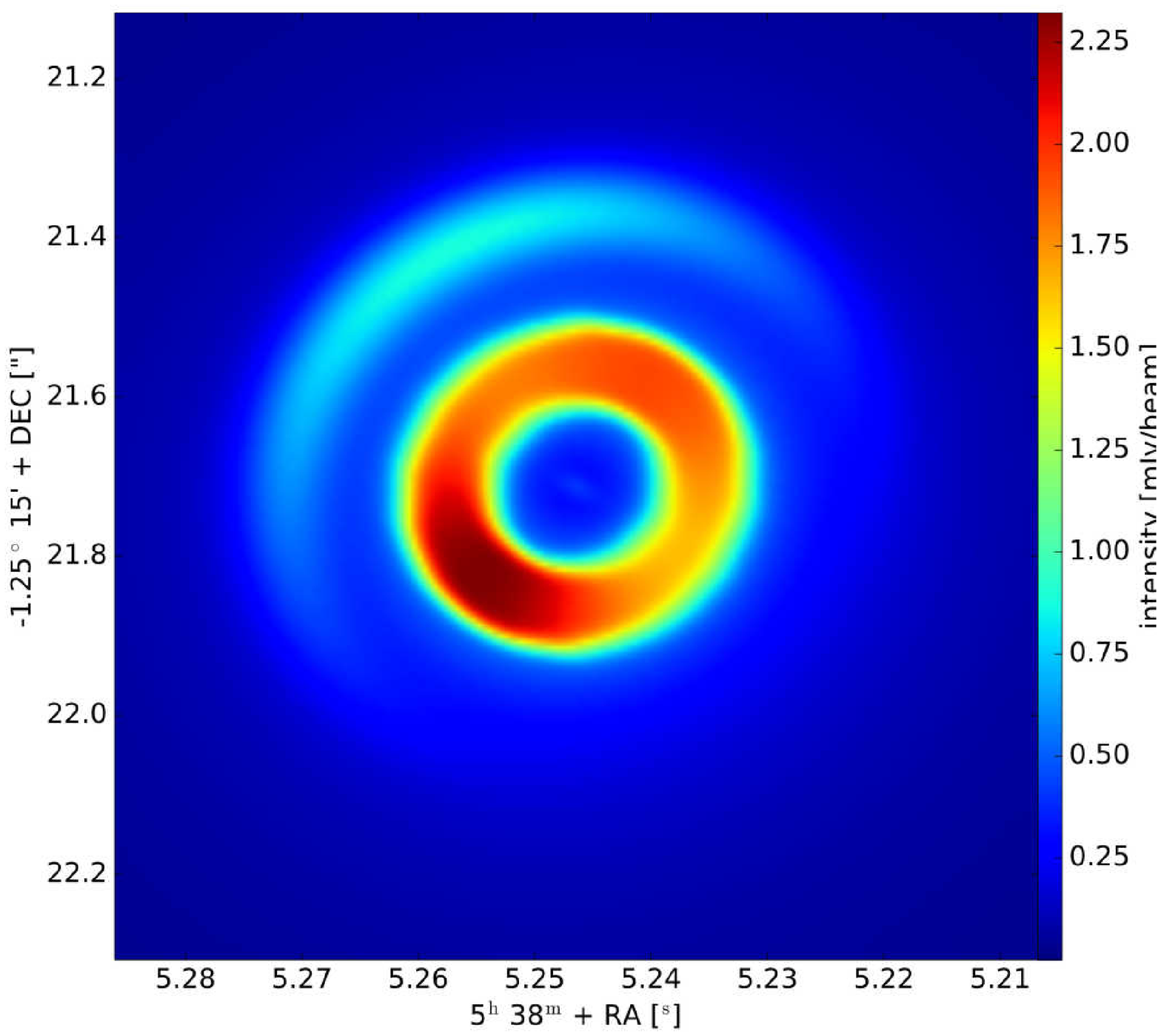} \\
     \includegraphics[angle=0,scale=0.27]{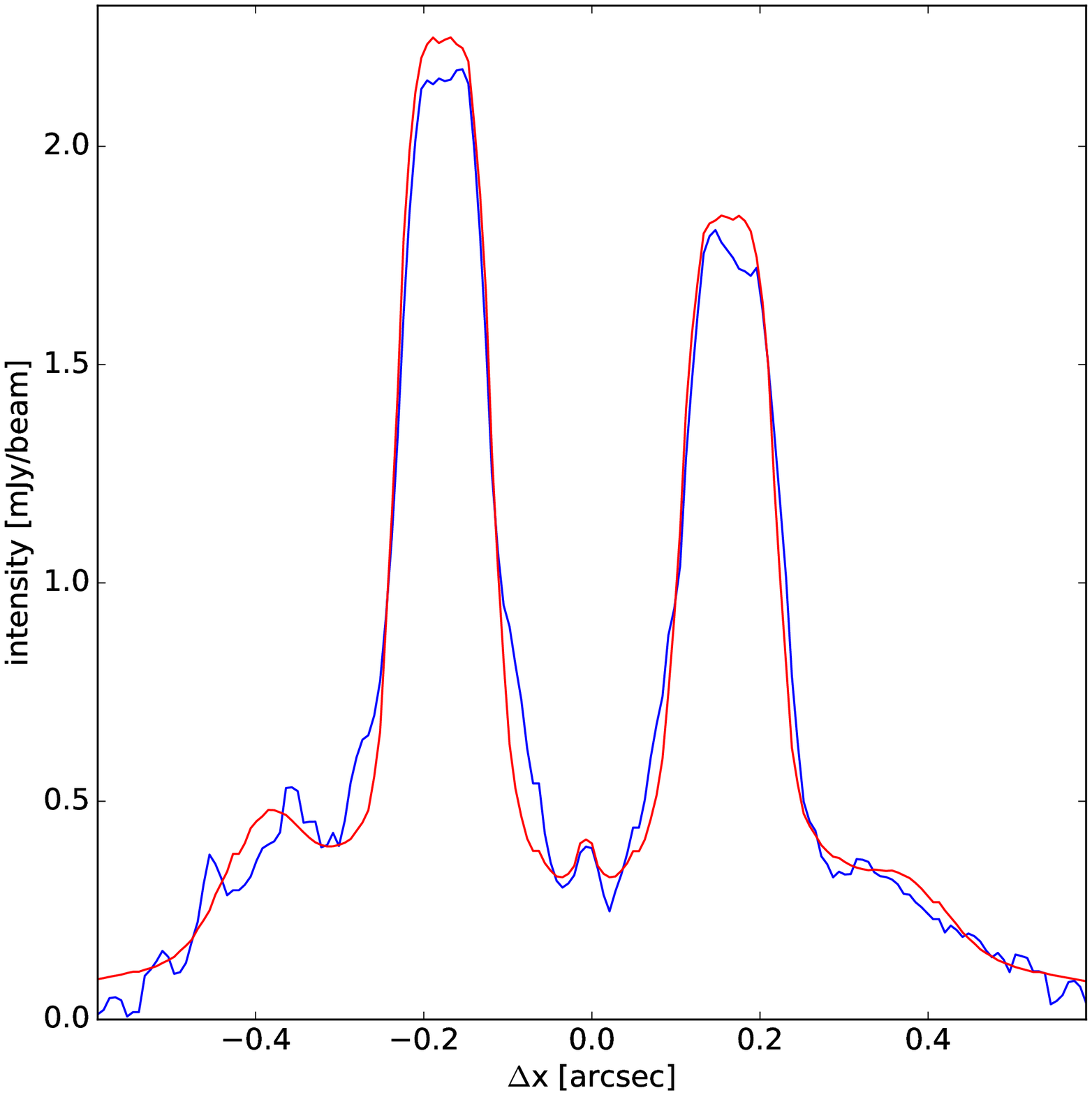} &
     \includegraphics[angle=0,scale=0.27]{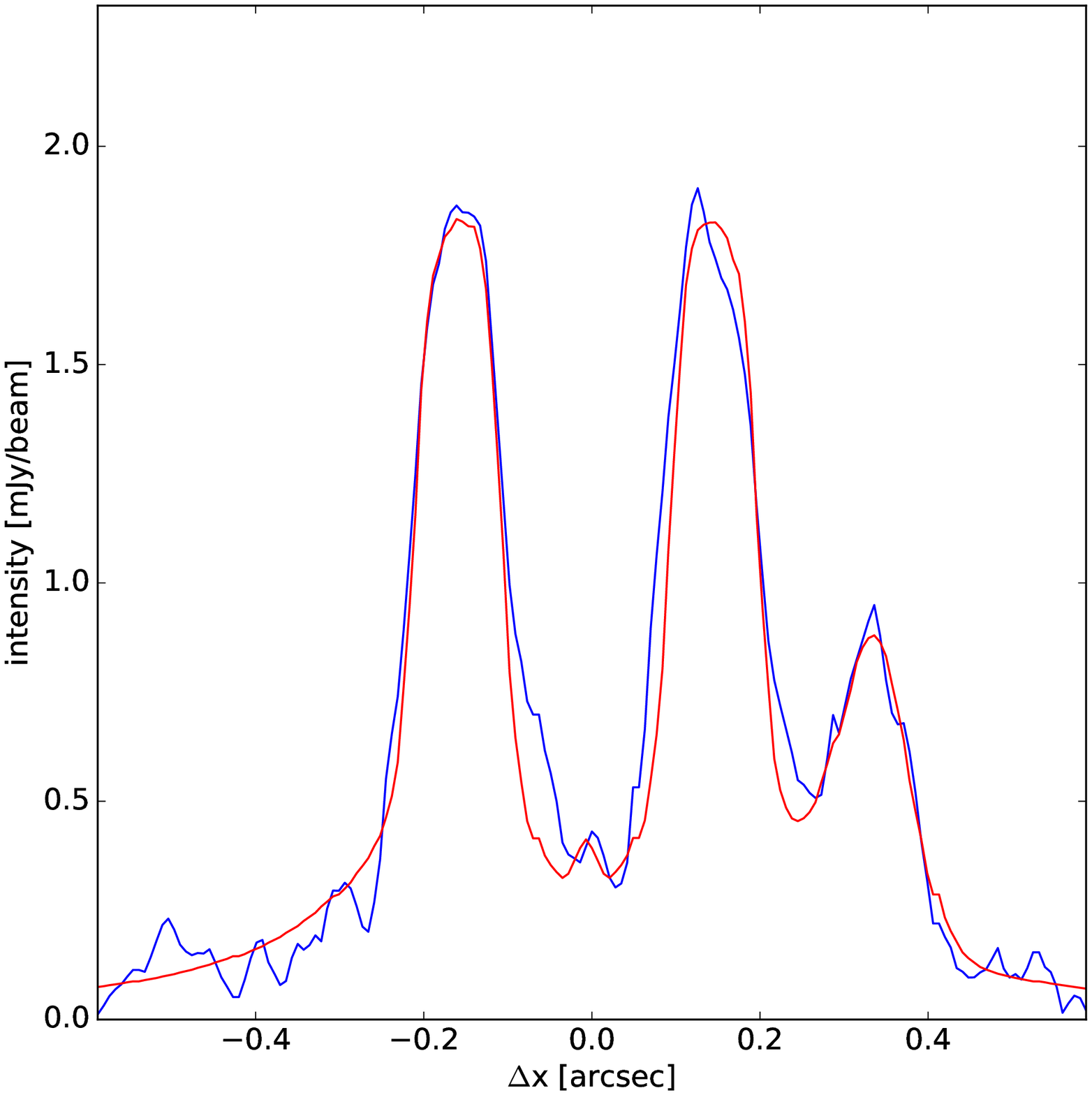}
  \end{array}$
  \caption{
    Top: ALMA 870\,$\mu$m image with the spatial components labeled (left), and 
    overplotted with the $H$-band scattered light image by \citet[][middle, contours at 1, 2, 3, 4, 5\% of peak intensity]{oht16}
    and the NIRC2 $L'$-band aperture masking image by \citet[][right, contours at 2, 3, 4, 5, 20, 40, 60, 80\%]{wil17}.
    Middle: ALMA image (left) and synthetic image for our best-fit model (right).
    Bottom: Radial intensity cuts along the disk major (left) and minor axis (right), for the ALMA image (blue line) and the model (red line).
  \label{fig:model}
  }
\end{figure*}

The 870\,$\mu$m continuum image (Fig.~\ref{fig:model}) reveals the following spatial components: 
\begin{itemize}
\item[(a)] an unresolved, central emission peak near the location of the star,
\item[(b)] a ring of emission (radii $\sim0.15{...}0.25\arcsec$) with significant substructure, namely strong excess flux in the south-eastern direction (R1) and weaker excess in the north-western direction (R2),
\item[(c)] a crescent-shaped structure (radii $\sim0.3{...}0.4\arcsec$) that extends along PA $\sim -40{...}+80^{\circ}$, but break down into two components (C1, C2), and 
\item[(d)] extended flux with a Gaussian half-width-at-half-maximum (HWHM) of $\sim0.3\arcsec$.
\end{itemize}

In order we quantify these components we fitted a geometric model consisting of the
following components (where $\theta$ denotes the azimuthal angle, counted eastwards of North): 
\begin{itemize}
\item[(1)] a {Uniform Disk with radius} $r_{\mathrm{CP}}$ and integrated flux $f_{\mathrm{CP}}$ {that represents} the central emission peak.
\item[(2)] a ring with {inner (outer) radius $r_{\mathrm{R}}^{\mathrm{in}}$ ($r_{\mathrm{R}}^{\mathrm{out}}$)} and integrated flux $f_{\mathrm{R}}$. Given the asymmetries {evident} in the ALMA image (features R1+R2), we give the fitting algorithm the flexibility to add two asymmetric components, whose azimuthal modulations we parameterise as $f_{\mathrm{R1}}(\theta)=(1-a_{\mathrm{R1}}\sin(\theta-\theta_{\mathrm{R1}}))^{\gamma_{\mathrm{R1}}}$ and $f_{\mathrm{R2}}(\theta)=(1-a_{\mathrm{R2}}\sin(\theta-\theta_{\mathrm{R2}}))^{\gamma_{\mathrm{R2}}}$.
\item[(3)] a ring with an azimuthal modulation to fit the crescent structure. The ring is parameterized with radius $r_{\mathrm{C}}$, Gaussian ring HWHM $\Theta_{\mathrm{C}}$, and the azimuthal modulation $f_{\mathrm{C}}(\theta)=(1-a_{\mathrm{C}}\sin(\theta-\theta_{\mathrm{C}}))^{\gamma_{\mathrm{C}}}$.  This component contributes $f_{\mathrm{C}}$ integrated flux.
\item[(4)] a {disk that represents} the extended flux component, contributing $f_{\mathrm{ext}}$ to the total flux.  
{The intensity profile is parameterised to connect smoothly with the ring component and drops with a radial power-law exponent $q_{\mathrm{ext}}$ inside and outside of the ring:
\begin{eqnarray}
I(r>r_{\mathrm{R}}^{\mathrm{out}})&=&\left(\frac{r}{r_{\mathrm{R}}^{\mathrm{out}}}\right)^{-q_{\mathrm{ext}}}\\
I(r_{\mathrm{R}}^{\mathrm{in}}{\leq}r{\leq}r_{\mathrm{R}}^{\mathrm{out}})&=&1\\
I(r<r_{\mathrm{R}}^{\mathrm{in}})&=&a\left(\frac{r}{r_{\mathrm{R}}^{\mathrm{in}}}\right)^{q_{\mathrm{ext}}}+b\\
b&=&\left(\frac{r_{\mathrm{R}}^{\mathrm{in}}+r_{\mathrm{R}}^{\mathrm{out}}}{r_{\mathrm{R}}^{\mathrm{out}}}\right)^{-q_{\mathrm{ext}}}\\
a&=&\left(\frac{1}{1-b}\right)^{-1/q_{\mathrm{ext}}}
\end{eqnarray}}
\end{itemize} 

All components are projected to mimic inclination effects (with inclination angle $i_{\mathrm{proj}}$ and disk minor-axis along PA $\theta_{\mathrm{proj}}$)
and the model images are convolved with the interferometric beam.
We explore the parameter space with a Differential Evolution optimisation algorithm {\citep{sto97}} and report best-fit parameters in Table~\ref{tab:model}.

\begin{deluxetable*}{lccc}
\tabletypesize{\scriptsize}
\tablecolumns{4}
\tablewidth{0pc}
\tablecaption{Model-fitting results (Sect.~\ref{sec:resultscont})\label{tab:model}}
\startdata 
\hline
\hline
\multicolumn{4}{c}{\bf Central peak (Significance 2$\sigma$)}\\
Central peak, uniform disk radius   & $r_{\mathrm{CP}}$   &  [\arcsec]                            &  $\lesssim 0.003$ \\ 
Central peak, integrated flux           & $f_{\mathrm{CP}}$ & [mJy]                                      &  $0.10 \pm 0.01$ \\
\hline
\multicolumn{4}{c}{\bf Ring structure (Significance 27$\sigma$)}\\
Ring inclination                               & $i_{\mathrm{proj}}$           &  [$^{\circ}$]                   &  $30.0 \pm 1.0$ \\ 
Ring PA                                             & $\theta_{\mathrm{proj}}$ &  [$^{\circ}$]                   &  $25.4 \pm 0.4$ \\ 
Ring inner radius                             & $r_{\mathrm{R}}^{\mathrm{in}}$ &  [\arcsec]             &  $0.129 \pm 0.001$ \\
Ring outer radius                            & $r_{\mathrm{R}}^{\mathrm{out}}$ &  [\arcsec]            &  $0.203 \pm 0.003$ \\
Ring flux                                           & $f_{\mathrm{R}}$ & [mJy]                                       &  $86.8 \pm 6.5$ \\  
Ring asymmetry R1, PA                     & $\theta_{\mathrm{R1}}$ &  [$^{\circ}$]         &  $105.5 \pm 1.6$ \\
Ring asymmetry R1, amplitude        & $a_{\mathrm{R1}}$ &                                     &  $0.154 \pm 0.011$ \\
Ring asymmetry R1, exponent        & $\gamma_{\mathrm{R1}}$ &                          &  $19.7 \pm 0.6$ \\ 
Ring asymmetry R2, PA                     & $\theta_{\mathrm{R2}}$ &  [$^{\circ}$]         &  $321.2 \pm 2.3$ \\
Ring asymmetry R2, amplitude        & $a_{\mathrm{R2}}$ &                                    &  $0.070 \pm 0.029$ \\
Ring asymmetry R2, exponent        & $\gamma_{\mathrm{R2}}$ &                          &  $30.2 \pm 0.9$ \\ 
\hline
\multicolumn{4}{c}{\bf Crescent structure (Significance 10$\sigma$)}\\
Crescent radius                            & $r_{\mathrm{C}}$ &  [\arcsec]                             &  $0.398 \pm 0.001$ \\
Crescent HWHM                           & $\Theta_{\mathrm{C}}$ &  [\arcsec]                    &  $0.053 \pm 0.002$ \\
Crescent flux                                & $f_{\mathrm{C}}$ &  [mJy]                                   &  $39.2 \pm 2.3$ \\  
Crescent asymmetry, PA                   & $\theta_{\mathrm{C}}$ &  [$^{\circ}$]             &  $5.0 \pm 1.2$ \\  
Crescent asymmetry, amplitude       & $a_{\mathrm{C}}$             &                            &  $0.751 \pm 0.002$ \\   
Crescent asymmetry, stretch-factor & $\gamma_{\mathrm{C}}$ &                            &  $2.57 \pm 0.15$ \\  
\hline
\multicolumn{4}{c}{\bf Extended structure (Significance 5$\sigma$)}\\
Extended emission, power-law exponent   & $q_{\mathrm{ext}}$    &                        &  $3.58 \pm 0.32$ \\
Extended emission flux                   & $f_{\mathrm{ext}}$           & [mJy]                        &  $173.9 \pm 23.4$    
\enddata 
\tablecomments{
We define inclination $0^{\circ}$ as face-on.  
PAs are measured East of North.}
\end{deluxetable*}

\begin{figure*}
  \centering
  $\begin{array}{cc}
     \includegraphics[angle=0, width=7.8cm]{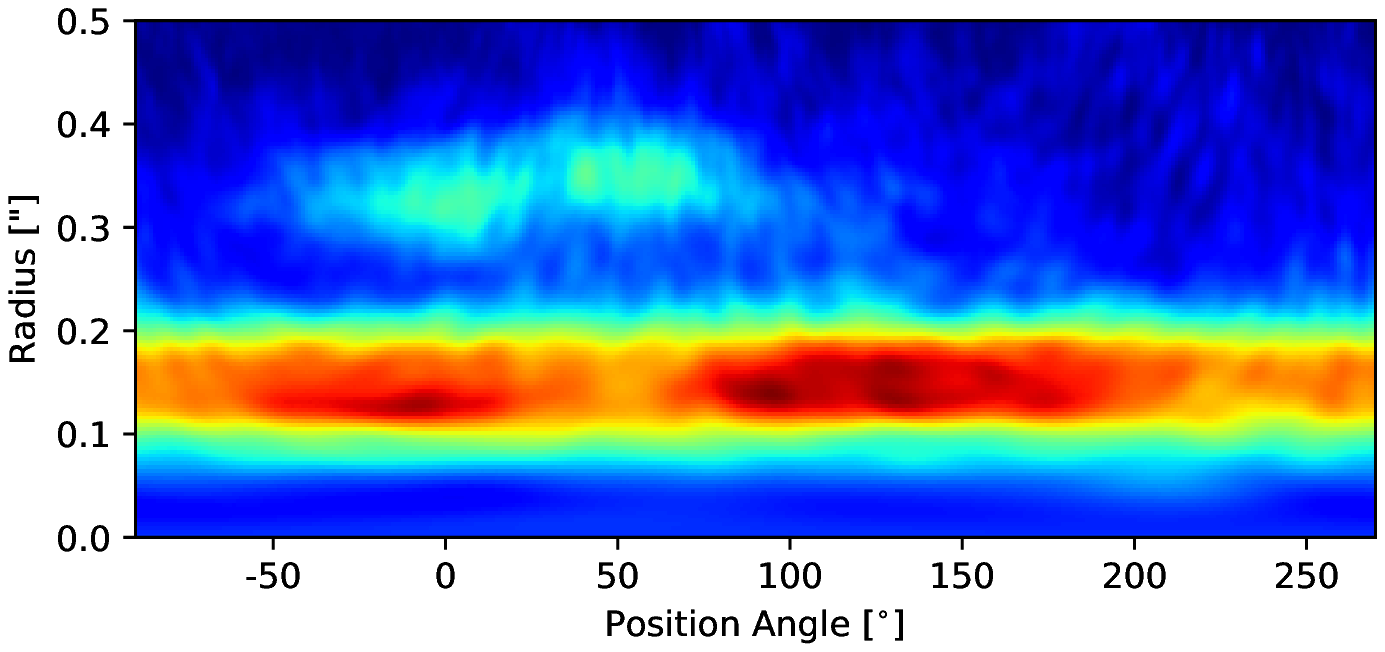} &
    \includegraphics[angle=0, width=7.8cm]{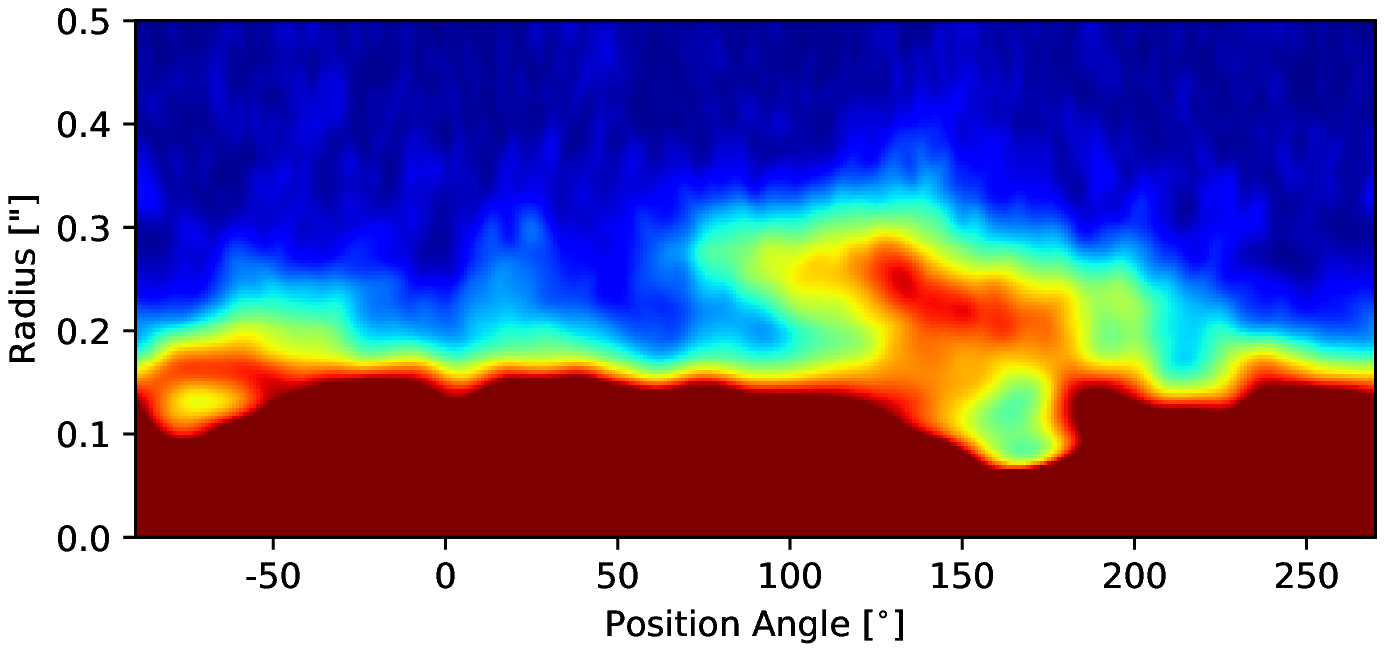}\\[5mm]
    \includegraphics[angle=0, scale=0.55]{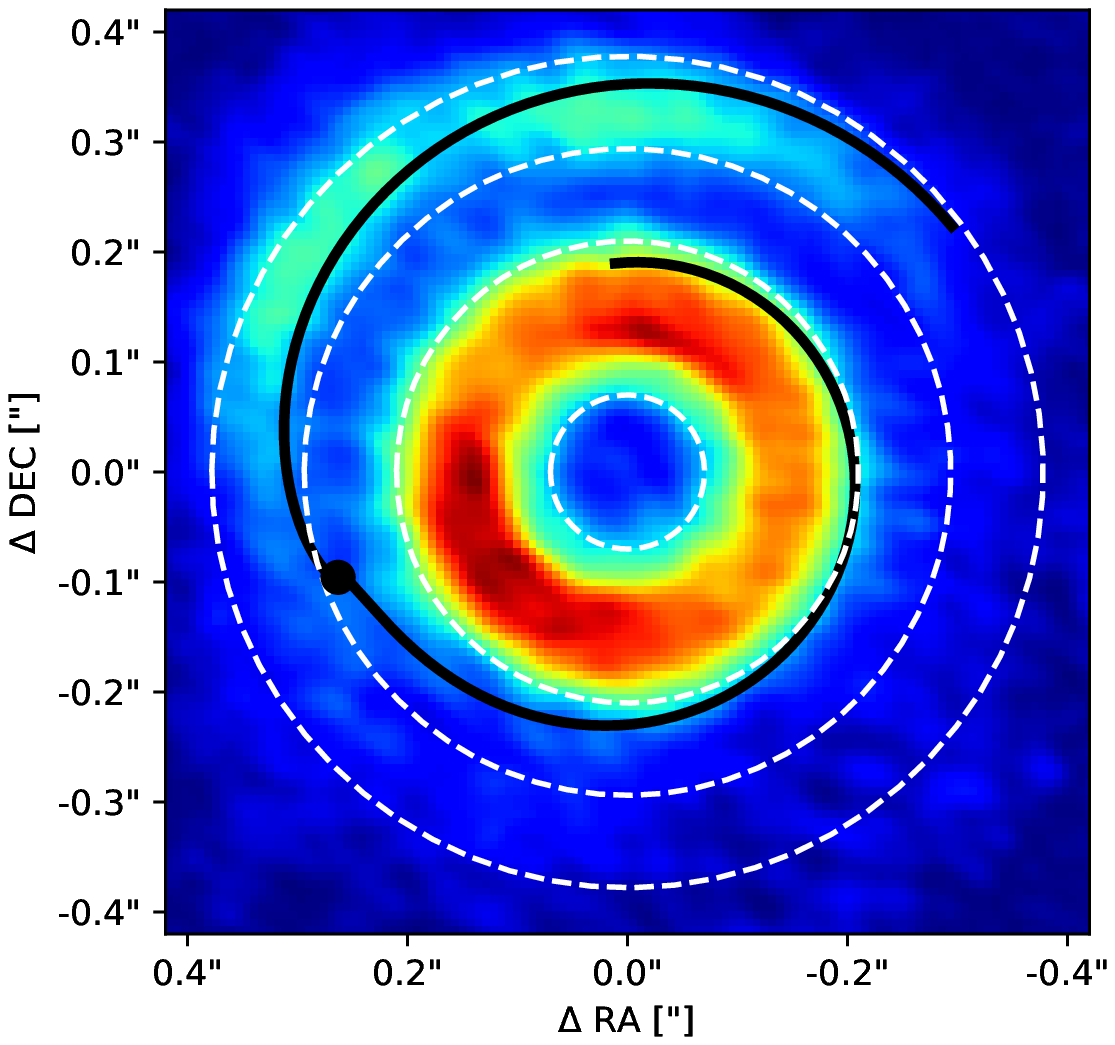} &
    \includegraphics[angle=0, width=6cm]{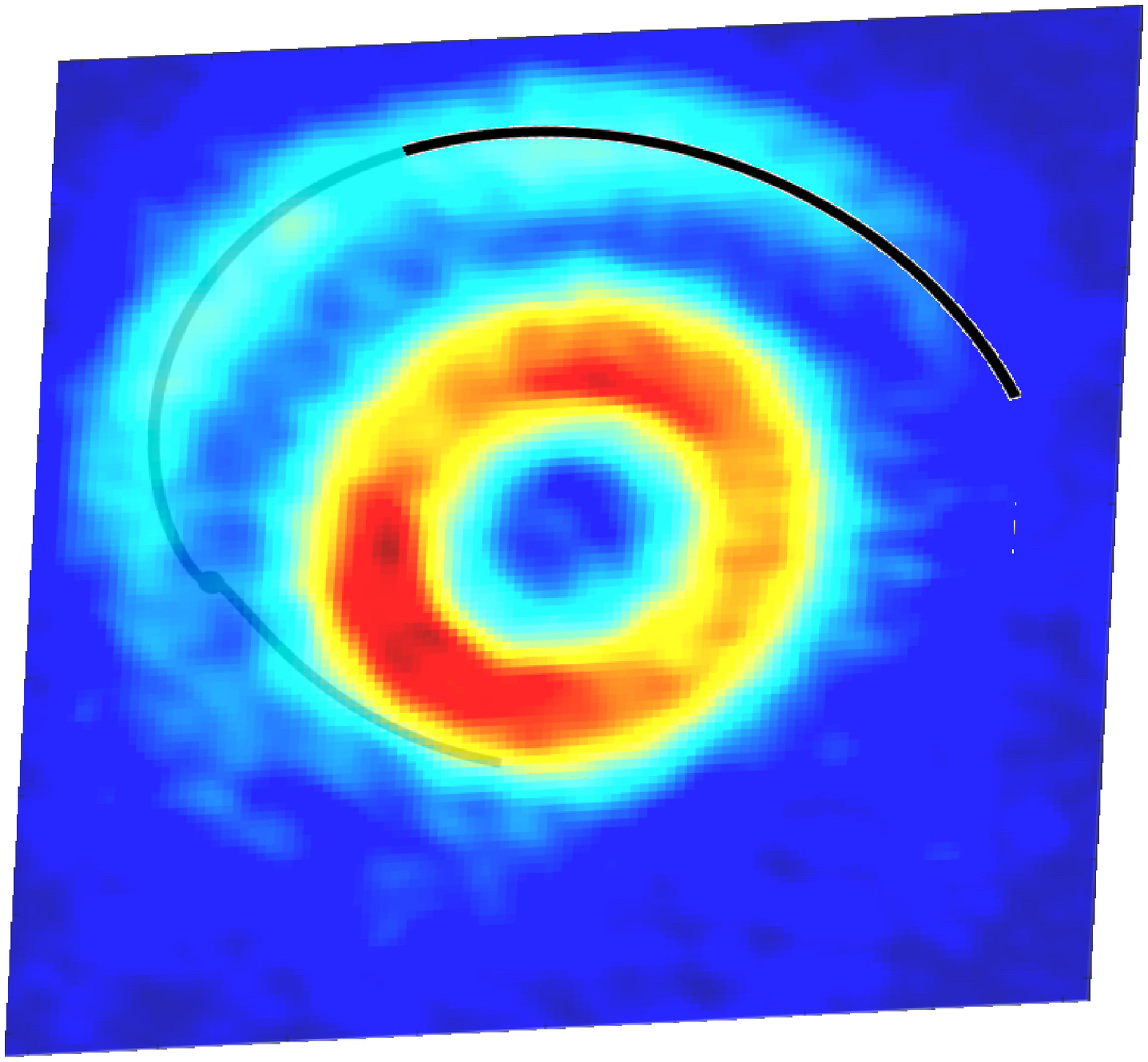}\\
     \multicolumn{2}{c}{\includegraphics[angle=0,width=11cm]{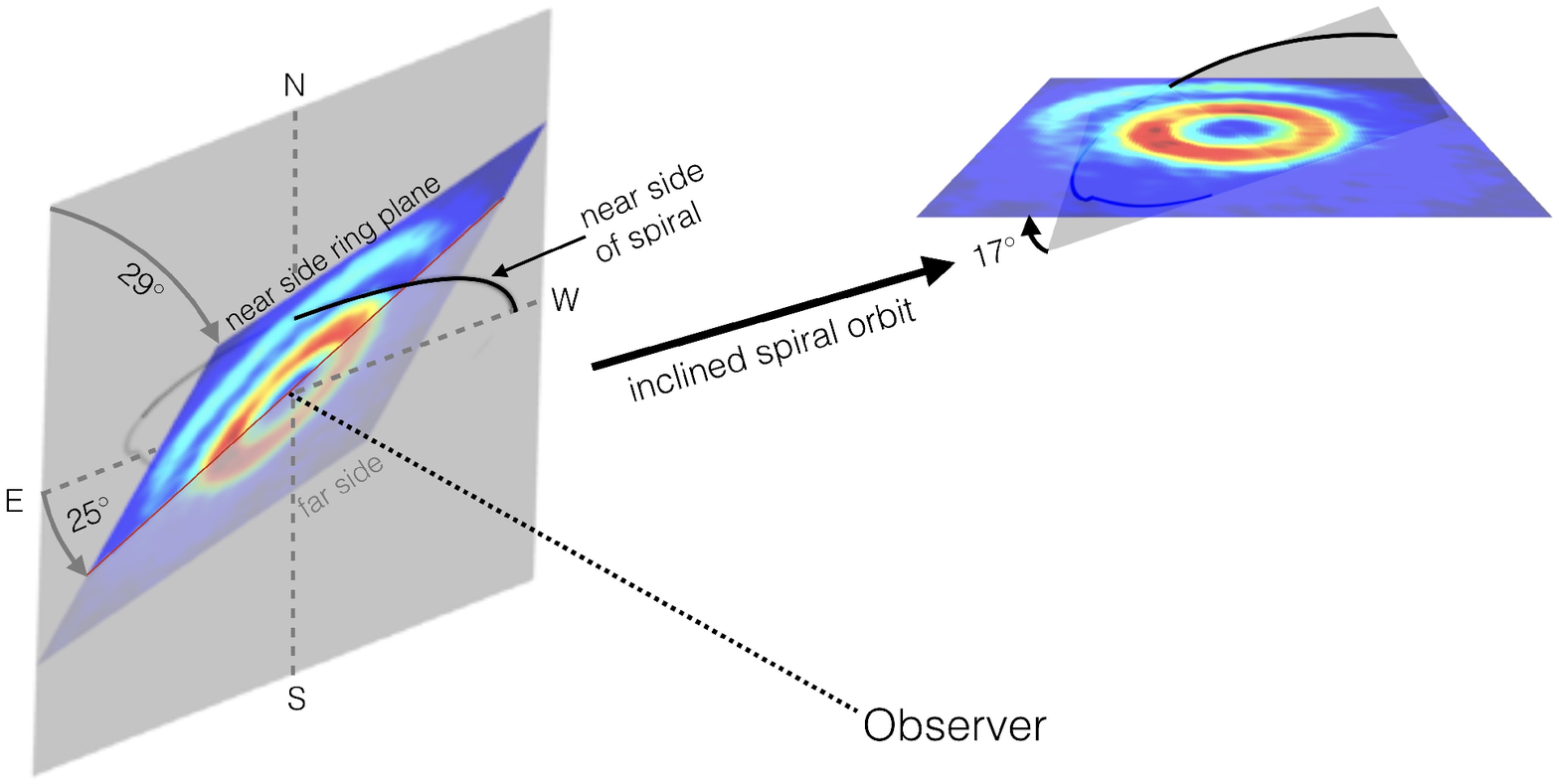}}
  \end{array}$
  \caption{
    Top: ALMA (left) and HiCIAO image (right) in polar projection.
    Middle-left: Deprojected ALMA image, with the analytic solution of a co-planar, 
    planet-triggered spiral arm (black curve; $r_p=90$\,au, $\theta_p=110^{\circ}$, $\epsilon=0.01$).
    The positon of the putative planet is marked as black dot.
    The dashed circles are included to guide the eye.
    Middle-right: Model with the spiral inclined by $17^{\circ}$ with respect to the disk plane.
    The part of the spiral located above the disk plane is indicated in black, 
    while the more distant part is grey.
    Bottom: Sketch of the viewing geometry in our inclined-spiral model (Sect.~\ref{sec:discussionplanet}).
  \label{fig:deprojected}
  }
\end{figure*}

\begin{figure*}
  \centering
  \vspace{2cm}
  \includegraphics[angle=0,scale=0.8]{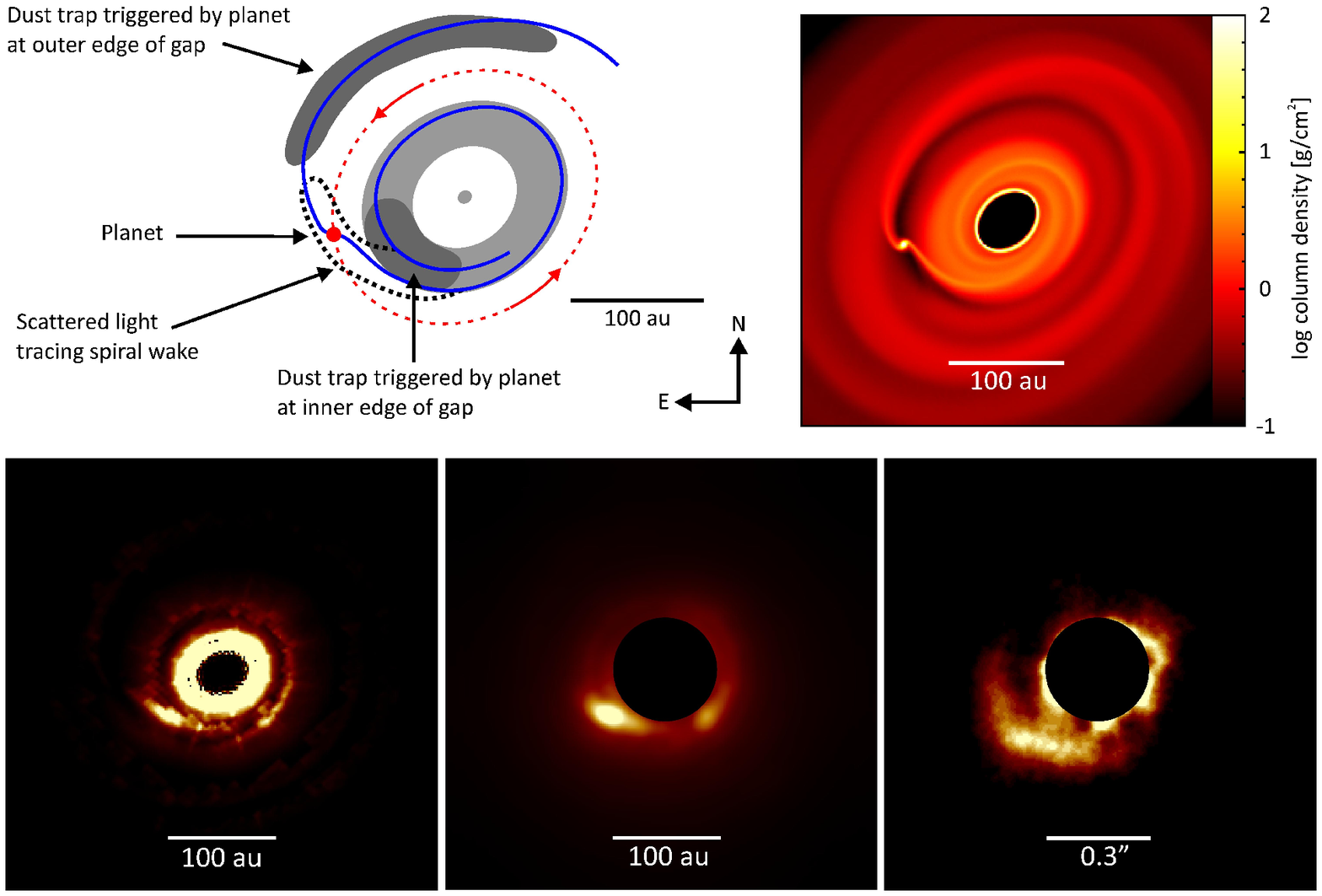} 
  \caption{
    Top-left: Illustration of the proposed physical scenario with a planet orbiting at $\sim100$\,au (Sect.~\ref{sec:discussionplanet}).
    Top-right: Gas column density of our radiation-hydrodynamic simulation (Sect.~\ref{sec:hydro}).
    Bottom: Synthetic 1.65\,$\mu$m scattered-light image at the full resolution (left) 
    and convolved with the HiCIAO point spread function (middle), and the observed HiCIAO image (right).
   \label{fig:hydro}
  }
\end{figure*}

\subsection{Line imaging}
\label{sec:resultslines}

\begin{figure*}
  \centering
   $\begin{array}{c@{\hspace{6.5mm}}c}
     \includegraphics[angle=0,scale=0.35]{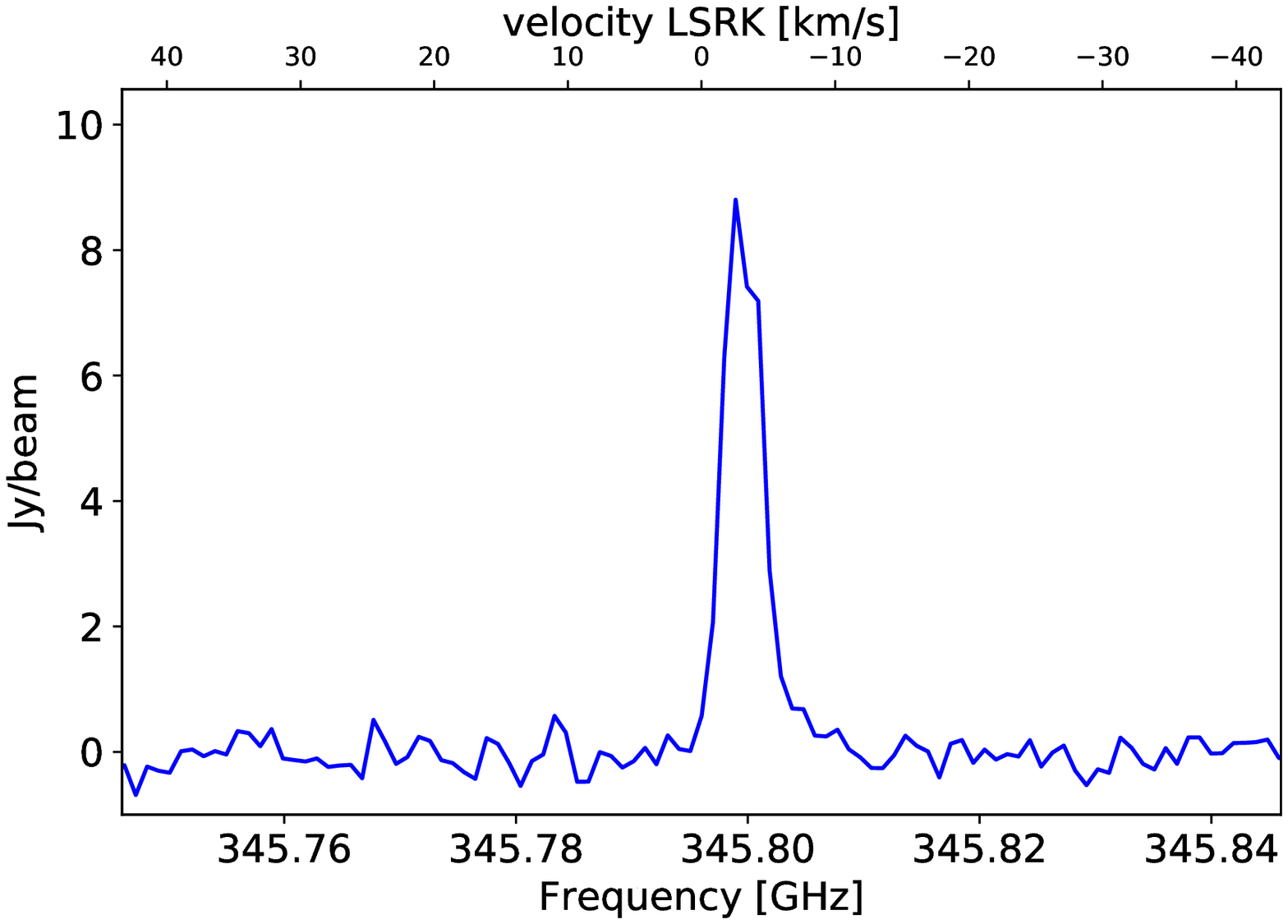} &
     \includegraphics[angle=0,scale=0.35]{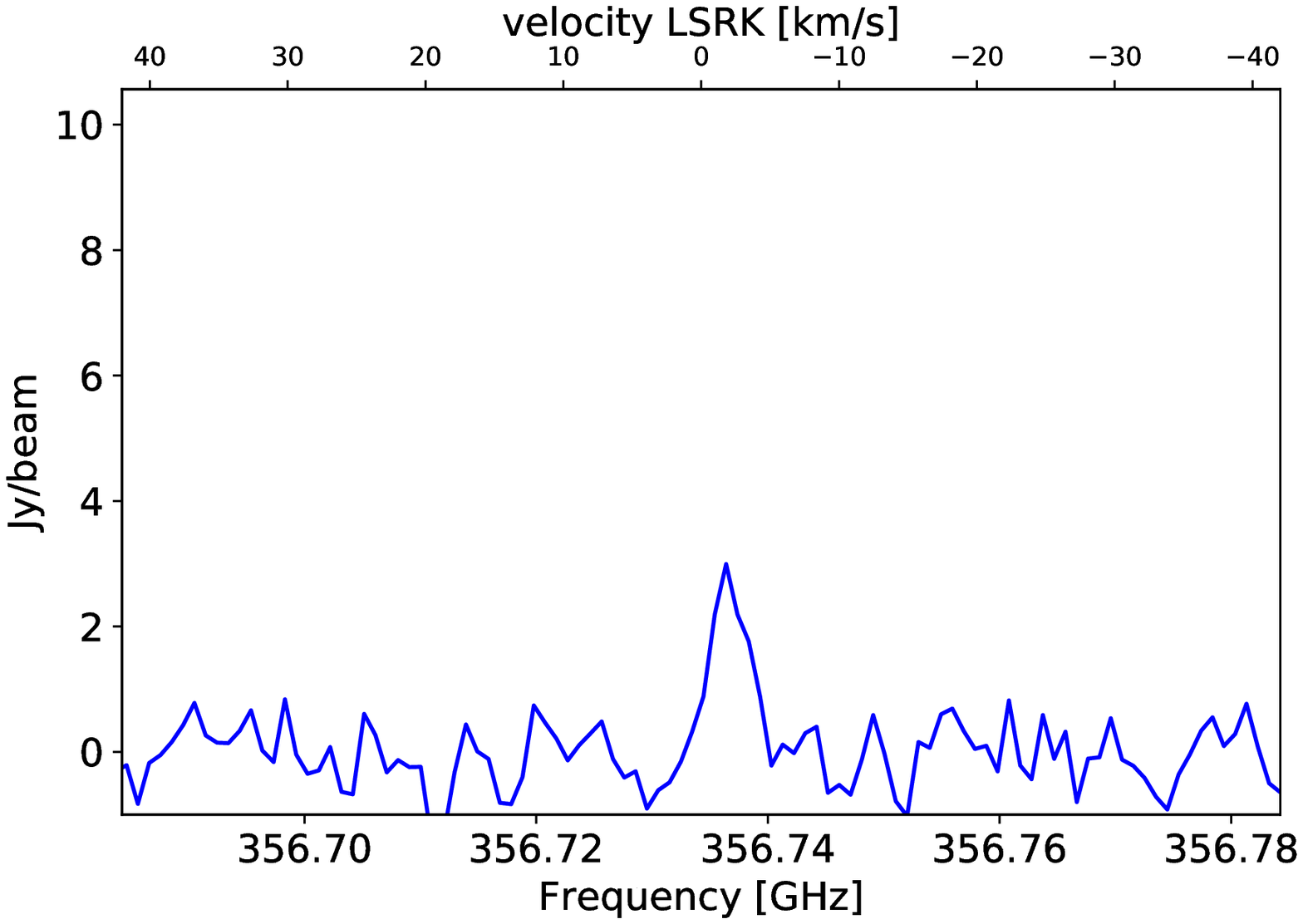}  \\[4mm]
     \includegraphics[angle=0,scale=0.24]{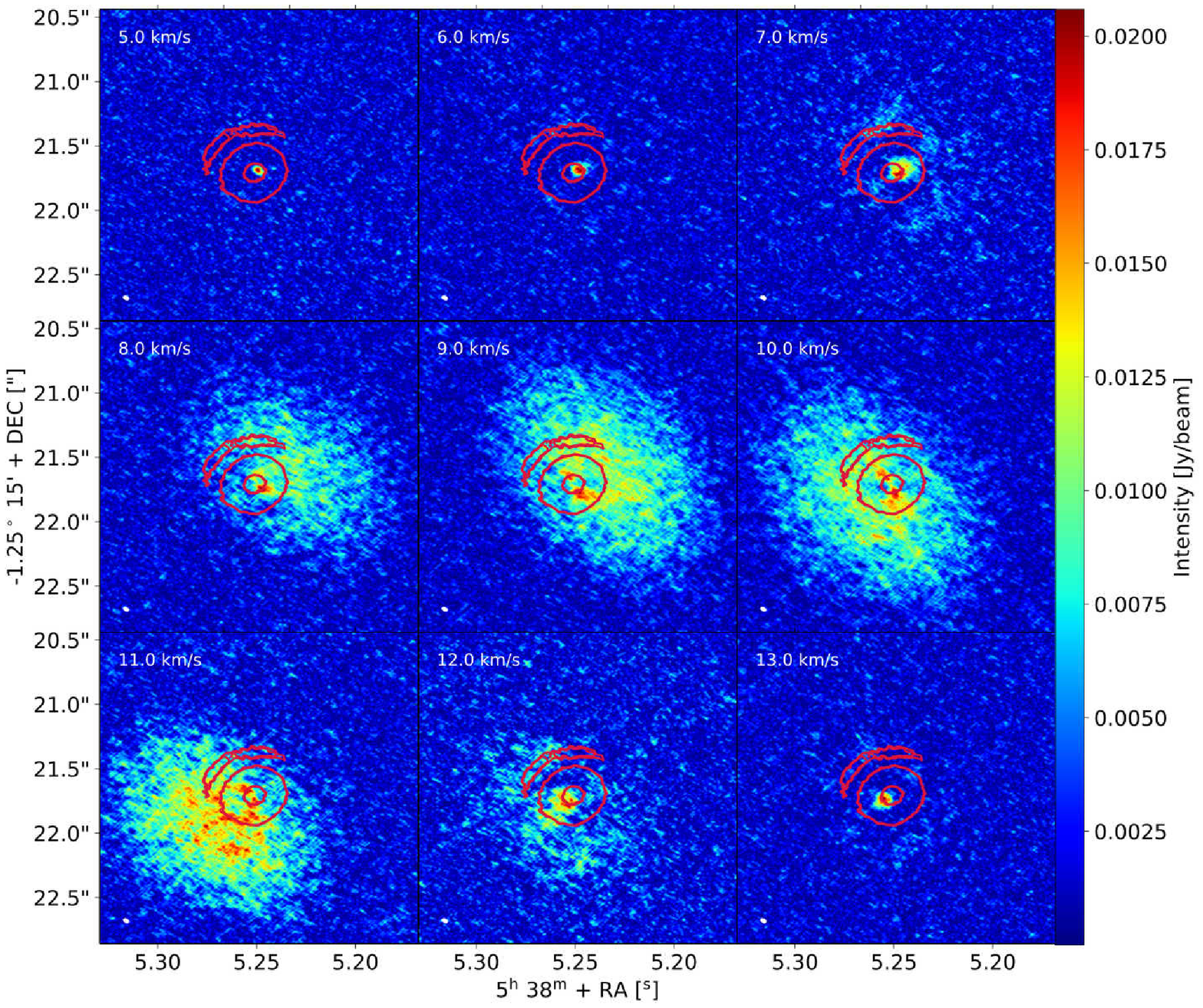} &
     \includegraphics[angle=0,scale=0.24]{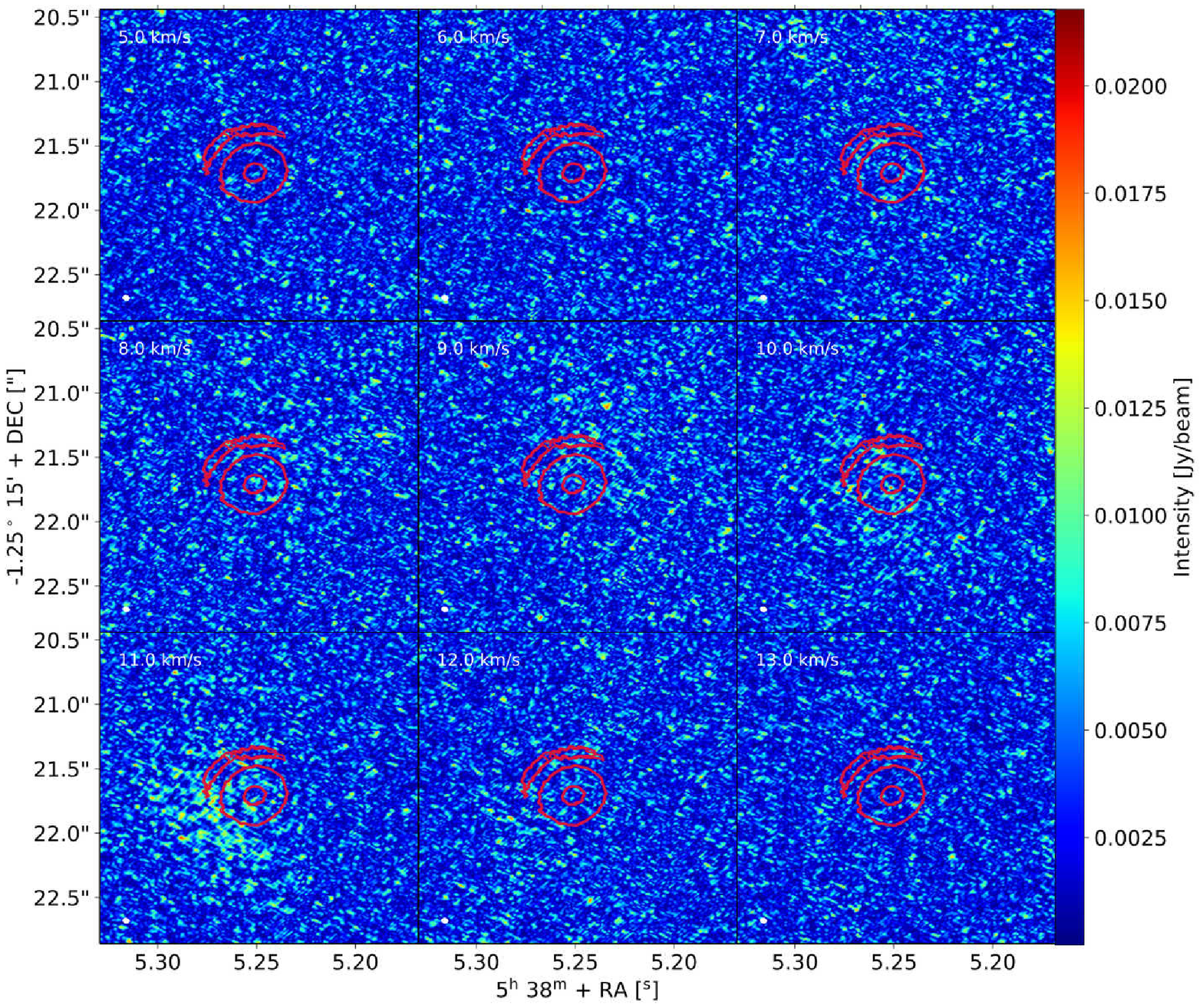}  \\[4mm]
     \includegraphics[angle=0,scale=0.24]{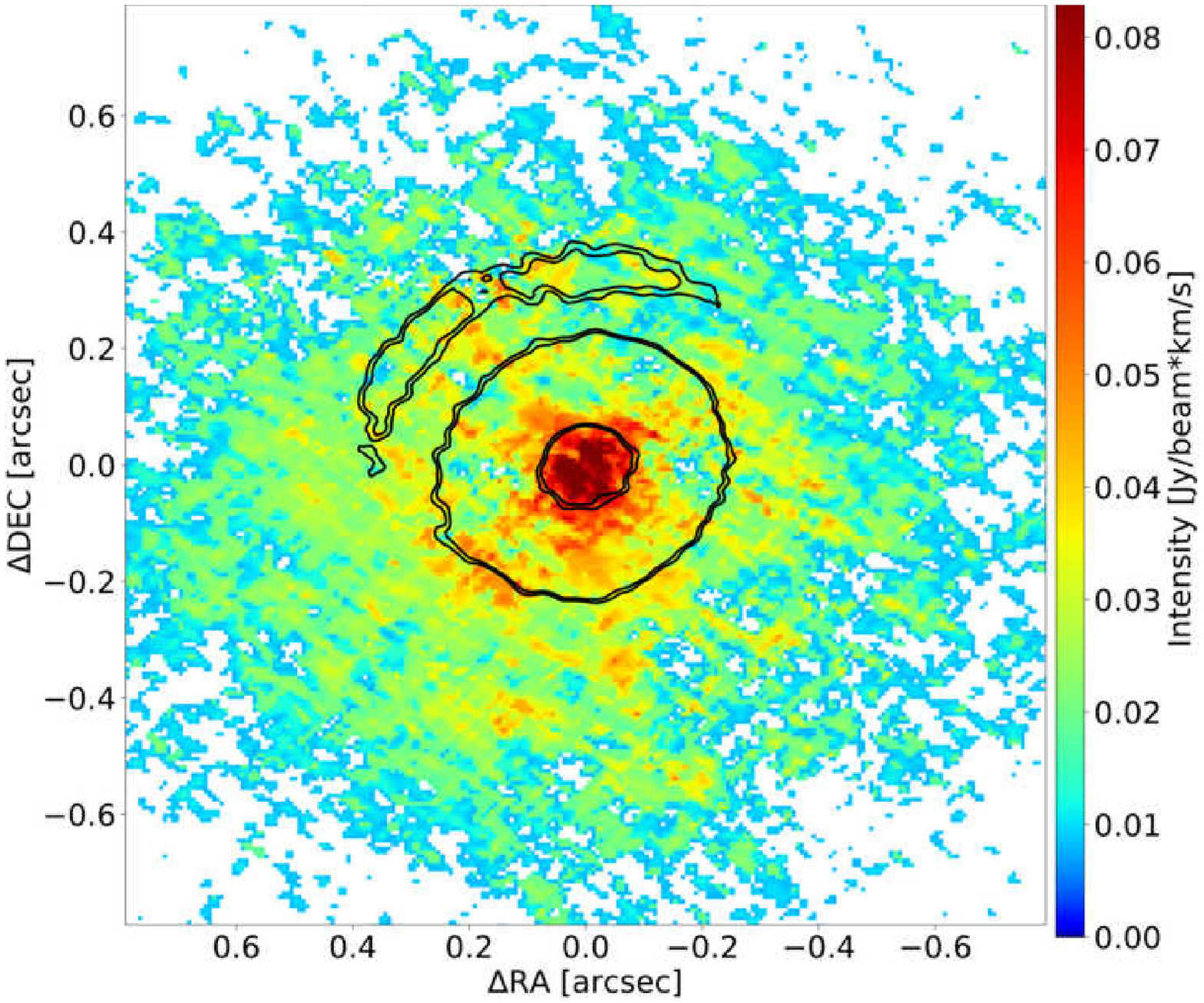} &
     \includegraphics[angle=0,scale=0.24]{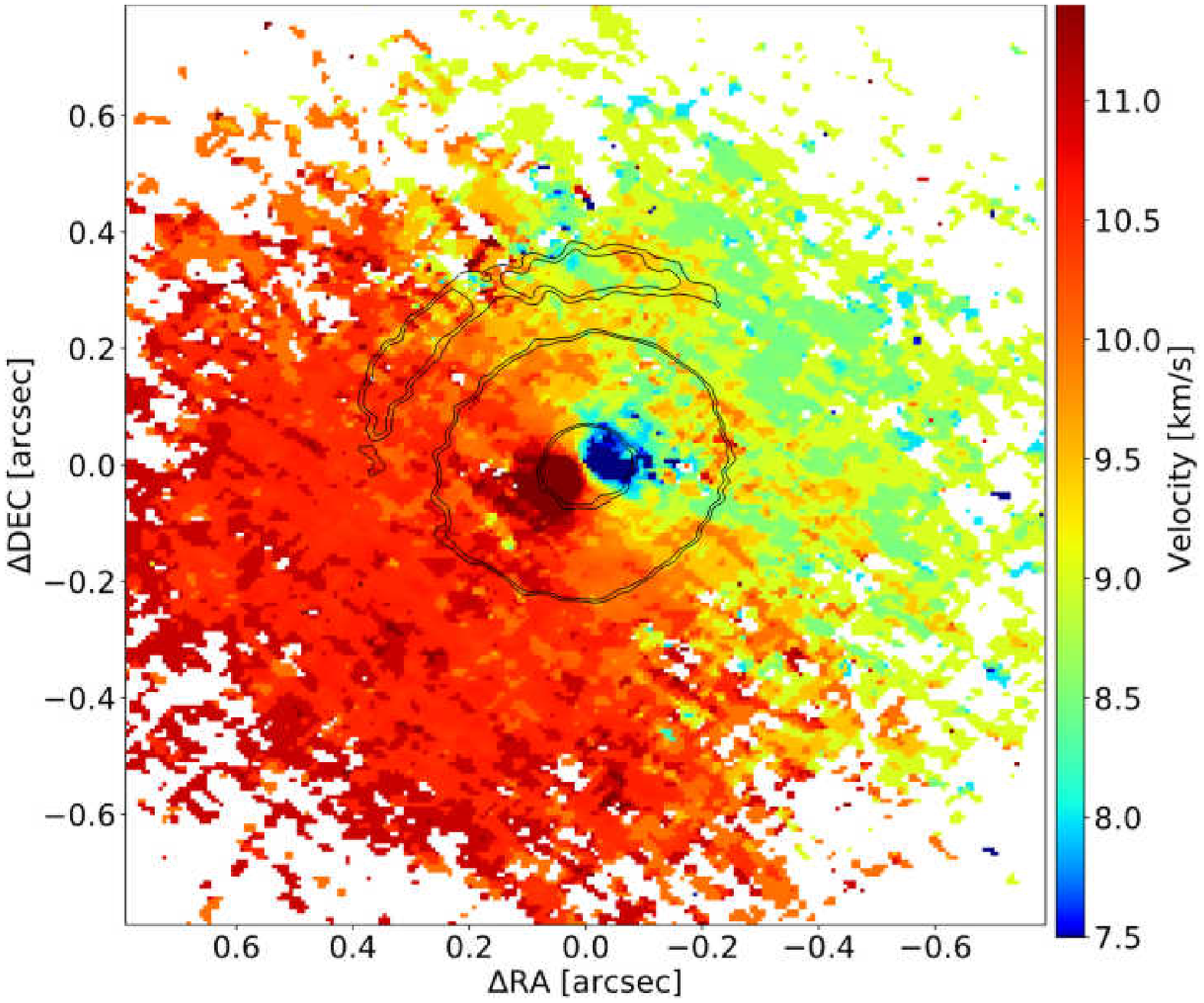}  \\
   \end{array}$
  \caption{
    Top: Spectra extracted from the ALMA data for CO\,(3-2) (left) and HCO$^{+}$\,(4-3) (right).
    Middle: Channel maps for the same lines.
    Bottom: CO moment~0 (left) and 1 (right) maps.
    The contours indicate the continuum-emitting structure.
    \label{fig:channelmaps}
  }
\end{figure*}

The ALMA spectrum (Fig.~\ref{fig:channelmaps}, top) exhibits line emission from CO\,(3-2) 
and H$^{12}$CO$^{+}$\,(4-3). Despite the relatively low signal-to-noise, we are able to draw
the following basic conclusions:

The CO channel maps and moment maps (Fig.~\ref{fig:channelmaps}, middle-left+bottom) show a rotation profile 
with the blue-shifted lobe to the North-West and the red-shifted lobe South-East of the star.
The CO flux drops off rapidly with increasing separation from the star, 
reflecting the radial temperature profile in the disk surface layer.
Interestingly, we see some excess of CO emission near the location of the 
crescent structure (C1-C2)
and in the south-eastern part of the disk near feature R1 outside of the ring structure.

The H$^{12}$CO$^{+}$ channel maps (Fig.~\ref{fig:channelmaps}, middle-right) suggest that the 
H$^{12}$CO$^{+}$-emitting gas also follows the disk rotation.
As with CO, we see some excess emission in the south-eastern part of the disk near feature R1.

\section{Discussion}
\label{sec:discussion}

\subsection{Multi-wavelength constraints on the disk structure}
\label{sec:discussiondiskstructure}

Some of the ALMA features can be directly associated with 
structures that have been previously resolved with VLTI NIR+MIR 
long-baseline interferometry \citep{kra13}:

The unresolved central peak seen by ALMA corresponds
likely to the optically-thick dust resolved with VLTI/AMBER NIR interferometry 
at the dust sublimation radius (0.15-0.22\,au).
The fact that the central peak is unresolved with ALMA confirms that this 
disk component is very narrow ($\lesssim4$\,au).

The MIR emission (8-13\,$\mu$m) was found to contain contributions
from an optically-thick disk at $38-70$\,au 
and emission from optically thin, carbonaceous dust grain located further in \citep{kra13}.
The size of this component matches with the ring-shaped emission 
detected in the ALMA image ($41-65$\,au).

The $L'$-band emission ($3.43-4.13\,\mu$m) was resolved 
with aperture masking interferometry,
where \citet{wil17} produced an image that shows $5\%$ of the 
$L'$-band emission arranged in an arc-like structure North of the star.
Overlaying the aperture masking and ALMA image (Fig.~\ref{fig:model}, top-right)
shows that the extended $L'$-band flux {traces likely thermal emission from} 
the northern part of the ALMA ring.
In order to emit significantly at such large distances from the stars, these particles
likely are quantum-heated by the UV-radiation of the star \citep{kla17},
which is consistent with the strong PAH features in the spectrum \citep{kra13}
and the fact that the PAH emission is spatially extended \citep{kra15}.

\subsection{The crescent structure and ring asymmetry as dust traps}
\label{sec:discussiondusttrap}

We considered whether the sub-millimeter crescent might constitute the
over-density in a planet-induced spiral arm, or, alternatively, a dust trap.

In order to test the spiral-arm hypothesis, we deproject the ALMA image using the 
ring inclination/PA and overplot the analytic description 
of a planet-triggered spiral-arm \citep{ogi02}, where the free parameters are the separation 
$r_p$ and PA $\theta_p$ of the planet and the dimension-less parameter $\epsilon$ that 
is proportional to the sound speed.
Figure~\ref{fig:deprojected} (middle-left) shows that the crescent does not follow the pitch angle of
the spiral-arm model and seems to be co-radial, instead.
We note that the agreement between the spiral-arm model and the crescent morphology
can be improved if the spiral is inclined with respect to the disk plane,
as illustrated in Fig.~\ref{fig:deprojected} (middle-right + bottom), where we adopted a misalignment angle of $17^{\circ}$.
The projection of the inclined spiral provides a better fit to the 
shape of the crescent, with a lower pitch angle than for a co-planar spiral. 
Interestingly, this scenario provides also a natural explanation for the brightness drop that is observed
along PA $25^{\circ}$ (between C1 and C2; Fig.~\ref{fig:model}).
In our inclined-spiral model, the disk plane and spiral plane cross along this PA
and the shadow casted by the geometrically-thin, optically-thick inner disk might lower the 
temperature at this location in the spiral, potentially explaining the observed sub-millimeter brightness drop.
The misalignment might suggest that the orbit of the planetary perturber 
is inclined with respect to the disk plane.  Various mechanisms have been identified that
could have resulted in such a configuration, including the dynamical interaction with other
bodies in the system (such as the inferred body at 5\,au) or a warp in the primordial disk that formed the planet.

However, we think that a dust-trapping vortex provides the most likely explanation
for the observed co-radial structure of the crescent, its high azimuthal contrast and azimuthal extend.
In a dust trap, the particles get trapped in a pressure bump and converge azimuthally towards the pressure maximum 
\citep{bra08}.  This allows them to grow efficiently to millimeter-size, which causes the region
to appear prominently at sub-millimeter wavelengths. 
Several hydrodynamic instabilities have been proposed that could trigger dust-trapping vortices, including 
the Rossby Wave Instability \citep{lov99},
the baroclinic instability \citep{kla03}, 
and the Kelvin-Helmholtz instability \citep{lit07}.
Interestingly, the Rossby Wave Instability is predicted to trigger the formation
of vortices not only at the pressure gradient at the outer edge of a
planet-cleared gap, but also near the inner edge \citep{li05}.
This leads us to suggest that the asymmetry R1 might also represent 
a vortex that formed at the inner edge of the gap.
The observed 30\% azimuthal modulation might trace the
mm-sized dust population that is superposed on the higher-density
population of small grains traced by our earlier MIR interferometry observations
\citep{kra13}.  The small grains are well-mixed with the gas and would appear
as a symmetric ring.

The high resolution offered by our observation enables us
to measure the vortex elongation (i.e.\ the ratio between the radial width and azimuthal extent), 
which is an important quantity for comparison with theoretical models.
We estimate elongations of 3.1 and 5.5 for the  ``ring'' and ``crescent'' vortex, respectively.
\citet{les09} predicted a stable regime for vortices with aspect-ratios between 4 and 5.9, 
although these numbers were computed for the gas and are thus not suitable for
direct comparison with our observed values.

\subsection{Planet-triggered dust traps and a spiral wake} 
\label{sec:discussionplanet}

The key observational diagnostics that we derive for V1247\,Ori include
a ring-like feature that is separated from an asymmetric disk segment by a gap
with a density depletion of at least one order-of-magnitude (likely more if one assumes 
that the dust grains are larger in the crescent).  
The crescent (C1+C2) and the ring (R1) appear to be smoothly 
connected through a spiral-arm-like scattered-light feature (S1).  
This scattered-light feature coincides with the slight density excess in the gap (``bridge'').

In order to explain these intriguing characteristics, we propose a scenario,
where a planet orbits at a separation of $\sim100$\,au={$0.3$\arcsec} 
from the star and clears the gap between the ring and the crescent structure (Fig.~\ref{fig:hydro}, top-left).
The dynamical influence of this planet could trigger dust traps on both
edges of the gap, namely in the crescent (C1+C2) and possibly also in the ring (R1).
The accretion streamers that feed the planet exhibit an increased vertical scale-height, 
which manifests in the observed spiral arm seen in scattered light.
The small dust grains coupled with the gas in the accretion streamers could cause the local enhancement
in dust density that is observed in the ``bridge'' connecting the ring and crescent.
This scenario is consistent with the disk-rotation direction
determined by \citet{wil17}, who determined the 3-dimensional disk orientation
by combining information on the rotation direction of the disk (from the CO kinematics) 
with the direction of motion of an orbiting structure detected at $\sim5$\,au. 
They conclude that the northern disk part is facing towards the observer.
Adopting this direction of motion (indicated with arrows in Fig.~\ref{fig:hydro}), 
we find that the spiral-arm is tailing, as expected for planet-triggered spiral wakes.

The scale-height increase in the spiral might also be responsible for the
$^{12}$CO and H$^{12}$CO$^{+}$ excess line emission {$\sim0.2{...}0.3$\arcsec} South-East of the star 
(Sect.~\ref{sec:resultslines}; Fig.~\ref{fig:channelmaps}).
For optically-thick lines, like $^{12}$CO, the surface brightness is proportional to the gas temperature.
{In the spiral structure (S1), the increased scale height will allow more efficient heating of the surface layer gas,
resulting in the excess line emission from this region.}

Besides the aforemention scenario with a planet at $\sim100$\,au, we also 
considered an alternative scenario, 
where a planet orbits at $\lesssim30$\,au and clears the region inside the ALMA ring. 
The planet generates a primary vortex (which dissipates and evolves into the 
ring observed in the ALMA image) and triggers the formation of a second-generation 
vortex further out, corresponding to the ALMA crescent.
This scenario predicts two distinct spiral arm structures observable in scattered light,
namely one spiral arm that is triggered by the crescent structure 
(imaged by HiCIAO) and one spiral arm triggered by the planet itself 
(yet undetected, but possibly located inside the ring structure).
In this scenario, the planet might already have been detected at 5-6\,au with the 
sparse aperture masking observations by \citet{wil17}.
This scenario has been proposed by \citet{van16} to explain the 
ALMA ring+crescent morphology in \object{HD135344B},
where indeed two spiral arms have been imaged in scattered light. 
However, there are significant differences with respect to the relative positions 
of the scattered light+ALMA {features:
In \object{HD135344B} the ALMA feature (their F1)
is cospatial with the end of the spiral arm (their S1), as expected if the 
spiral density wave is triggered by the vortex itself \citep{van16}.
This is not the case for V1247\,Ori, which is more consistent with
the hypothesis that both are directly triggered by a planet.}

\section{Radiation Hydrodynamics Simulation}
\label{sec:hydro}

We conducted dedicated smooth particle hydrodynamics
simulations with the specific parameters of V1247~Ori in order to test whether
the scale-height variation of a planet-induced spiral wake could
result in the detected scattered-light signature.
Our simulation adopts a fixed gas-to-dust ratio and a relatively high viscosity ($\alpha\approx0.01$)
and is therefore not suited to simulate the vortex formation itself, which will require further dedicated theoretical work.
Given the large number of free parameters involved, 
it is also out of the scope of this paper to model the images quantitatively.

Our simulations were performed using the code by \citet{ben90}, \citet{bat95} and \citet{pri07}
that was adopted for modelling embedded protoplanets by \citet{ayl09,ayl10}. 
The disk extends over radii $r=30-300$\,au, with a constant scale height $h/r=0.05$ 
and a surface density $\Sigma{\propto}r^{-1.5}$,
corresponding to a disk mass of 0.013\,M$_{\sun}$. 
Our calculation was performed with a resolution of $10^{6}$ particles
and assuming a 3\,M$_{\mathrm{Jup}}$ planet mass. 
The planet was placed at $r=100$\,au and the disk was allowed to evolve. 
Snapshots were taken
once the simulation reached a quasi-steady state after 10 orbits (at the planet location).
We used the Monte-Carlo radiative transfer code TORUS \citep{har00} 
to create synthetic images, where we
assume a fixed dust-to-gas ratio of 100, a grain size distribution $a^{-3.5}$ for 
$0.005\,\mu\mathrm{m}\leq{a}\leq1\,\mu\mathrm{m}$, 
and silicate grain opacities \citep{dra84}. 

The synthetic scattered-light image (Fig.~\ref{fig:hydro}, bottom) shows a spiral-arm-like
structure that appears most prominently in the inwards-facing spiral arm triggered by the planet. 
This confirms that a planet-triggered spiral arm should be detectable in polarised light,
where the PA and pitch angle matches roughly the ones observed with HiCIAO.

Our simulation predicts a high column density in the vicinity of the planet (Fig.~\ref{fig:hydro}, top-right),
which opens the possibility that the circumplanetary disk might be detectable in sub-mm continuum emission.
This sub-mm flux is not detected, which might indicate that the sub-mm emissivity of the dust 
in the vicinity of the planet is lower than in other disk structures
\citep[e.g.\ due to dust filtration processes at the edge of the gap,][]{paa06}.
Sophisticated simulations with decoupled gas+dust populations and 
grain growth will be required to derive quantitative constraints.
Also, the planet emission might be confused with the bright emission from the nearby ring feature R1.

\section{Conclusions}
\label{sec:conclusions}

Our ALMA imaging of V1247\,Ori reveals a crescent structure whose 
morphology and properties match the predictions of a dust-trapping vortex.
The superb {0.04\arcsec}-resolution of our ALMA image reveals intriguing 
substructure in the dust trap and provides for the first time the opportunity to 
confront competing theories of vortex formation with detailed observational evidence.

We propose that the vortex is triggered by a planet orbiting at {$\sim100$\,au},
whose spiral wake is seen in scattered light {and likely in the} detected $^{12}$CO and H$^{12}$CO$^{+}$ line excess emission.
The continuum excess emission observed in the asymmetric ring {feature (R1)}
might constitute a second dust trap that formed near the inner edge of the
planet-cleared gap.

The protoplanet itself has not been discovered so far, neither in SPHERE H$\alpha$ spectral 
differential imaging \citep{wil17} nor in our ALMA imaging, 
possibly due to confusion with the bright ring structure.
Further efforts should be taken to achieve a direct detection, for instance with deeper ALMA imaging.

\acknowledgments

This paper makes use of ALMA data set ADS/JAO.ALMA\#2015.1.00986.S. 
We thank the German ALMA ARC for support, in particular Stefanie M\"uhle and Benjamin Magnelli.
Our team acknowledges support from the European Research Council (Grant Agreement Numbers 639889 and 339248), an
STFC Rutherford Fellowship/Grant (ST/J004030/1, ST/K003445/1),
Philip Leverhulme Prize (PLP-2013-110), 
NAOJ ALMA Scientific Research Grant (Number 2016-02A), and
NASA Exoplanet Research Program grants NNX16AJ75G and NNX17AF88G.
ALMA is a partnership of ESO (representing its member states), NSF (USA) and NINS (Japan), 
together with NRC (Canada), MOST and ASIAA (Taiwan), and KASI (Republic of Korea), 
in cooperation with the Republic of Chile. The Joint ALMA Observatory is operated by ESO, AUI/NRAO and NAOJ.
This work used the DiRAC Complexity system, operated by the University of Leicester IT Services, which forms part of the STFC DiRAC HPC Facility. 
This equipment is funded by BIS National E-Infrastructure capital grant ST/K000373/1 and STFC DiRAC Operations grant ST/K0003259/1.

{\it Facilities:} \facility{ALMA}


\begin{thebibliography}{31}
\expandafter\ifx\csname natexlab\endcsname\relax\def\natexlab#1{#1}\fi

\bibitem[{{Ayliffe} \& {Bate}(2009)}]{ayl09}
{Ayliffe}, B.~A., \& {Bate}, M.~R. 2009, \mnras, 393, 49

\bibitem[{{Ayliffe} \& {Bate}(2010)}]{ayl10}
---. 2010, \mnras, 408, 876

\bibitem[{{Bate} {et~al.}(1995){Bate}, {Bonnell}, \& {Price}}]{bat95}
{Bate}, M.~R., {Bonnell}, I.~A., \& {Price}, N.~M. 1995, \mnras, 277, 362

\bibitem[{{Benz}(1990)}]{ben90}
{Benz}, W. 1990, in Numerical Modelling of Nonlinear Stellar Pulsations
  Problems and Prospects, ed. J.~R. {Buchler}, 269

\bibitem[{{Birnstiel} {et~al.}(2013){Birnstiel}, {Dullemond}, \&
  {Pinilla}}]{bir13}
{Birnstiel}, T., {Dullemond}, C.~P., \& {Pinilla}, P. 2013, \aap, 550, L8

\bibitem[{{Brauer} {et~al.}(2008){Brauer}, {Dullemond}, \& {Henning}}]{bra08}
{Brauer}, F., {Dullemond}, C.~P., \& {Henning}, T. 2008, \aap, 480, 859

\bibitem[{{Dodson-Robinson} \& {Salyk}(2011)}]{dod11}
{Dodson-Robinson}, S.~E., \& {Salyk}, C. 2011, \apj, 738, 131

\bibitem[{{Draine} \& {Lee}(1984)}]{dra84}
{Draine}, B.~T., \& {Lee}, H.~M. 1984, \apj, 285, 89

\bibitem[{{Gaia Collaboration} {et~al.}(2016{\natexlab{a}}){Gaia
  Collaboration}, {Brown}, {Vallenari}, {Prusti}, {de Bruijne}, {Mignard},
  {Drimmel}, {Babusiaux}, {Bailer-Jones}, {Bastian}, \& et~al.}]{gaia1a}
{Gaia Collaboration} {et~al.} 2016{\natexlab{a}}, \aap, 595, A2

\bibitem[{{Gaia Collaboration} {et~al.}(2016{\natexlab{b}}){Gaia
  Collaboration}, {Prusti}, {de Bruijne}, {Brown}, {Vallenari}, {Babusiaux},
  {Bailer-Jones}, {Bastian}, {Biermann}, {Evans}, \& et~al.}]{gaia1b}
---. 2016{\natexlab{b}}, \aap, 595, A1

\bibitem[{{Harries}(2000)}]{har00}
{Harries}, T.~J. 2000, \mnras, 315, 722

\bibitem[{{Klahr} \& {Bodenheimer}(2003)}]{kla03}
{Klahr}, H.~H., \& {Bodenheimer}, P. 2003, \apj, 582, 869

\bibitem[{{Klarmann} {et~al.}(2017){Klarmann}, {Benisty}, {Min}, {Dominik},
  {Berger}, {Waters}, {Kluska}, {Lazareff}, \& {Le Bouquin}}]{kla17}
{Klarmann}, L., {et~al.} 2017, \aap, 599, A80

\bibitem[{{Kraus}(2015)}]{kra15}
{Kraus}, S. 2015, \apss, 357, 97

\bibitem[{{Kraus} {et~al.}(2013){Kraus}, {Ireland}, {Sitko}, {Monnier},
  {Calvet}, {Espaillat}, {Grady}, {Harries}, {H{\"o}nig}, {Russell},
  {Swearingen}, {Werren}, \& {Wilner}}]{kra13}
{Kraus}, S., {et~al.} 2013, \apj, 768, 80

\bibitem[{{Lesur} \& {Papaloizou}(2009)}]{les09}
{Lesur}, G., \& {Papaloizou}, J.~C.~B. 2009, \aap, 498, 1

\bibitem[{{Li} {et~al.}(2005){Li}, {Li}, {Koller}, {Wendroff}, {Liska},
  {Orban}, {Liang}, \& {Lin}}]{li05}
{Li}, H., {Li}, S., {Koller}, J., {Wendroff}, B.~B., {Liska}, R., {Orban},
  C.~M., {Liang}, E.~P.~T., \& {Lin}, D.~N.~C. 2005, \apj, 624, 1003

\bibitem[{{Lithwick}(2007)}]{lit07}
{Lithwick}, Y. 2007, \apj, 670, 789

\bibitem[{{Lovelace} {et~al.}(1999){Lovelace}, {Li}, {Colgate}, \&
  {Nelson}}]{lov99}
{Lovelace}, R.~V.~E., {Li}, H., {Colgate}, S.~A., \& {Nelson}, A.~F. 1999,
  \apj, 513, 805

\bibitem[{{Ogilvie} \& {Lubow}(2002)}]{ogi02}
{Ogilvie}, G.~I., \& {Lubow}, S.~H. 2002, \mnras, 330, 950

\bibitem[{{Ohta} {et~al.}(2016){Ohta}, {Fukagawa}, {Sitko}, {Muto}, {Kraus},
  {Grady}, {Wisniewski}, {Swearingen}, {Shibai}, {Sumi}, {Hashimoto}, {Kudo},
  {Kusakabe}, {Momose}, {Okamoto}, {Kotani}, {Takami}, {Currie}, {Thalmann},
  {Janson}, {Akiyama}, {Follette}, {Mayama}, {Abe}, {Brandner}, {Brandt},
  {Carson}, {Egner}, {Feldt}, {Goto}, {Guyon}, {Hayano}, {Hayashi}, {Hayashi},
  {Henning}, {Hodapp}, {Ishii}, {Iye}, {Kandori}, {Knapp}, {Kuzuhara}, {Kwon},
  {Matsuo}, {McElwain}, {Miyama}, {Morino}, {Moro-Mart{\'{\i}}n}, {Nishimura},
  {Pyo}, {Serabyn}, {Suenaga}, {Suto}, {Suzuki}, {Takahashi}, {Takami},
  {Takato}, {Terada}, {Tomono}, {Turner}, {Usuda}, {Watanabe}, {Yamada}, \&
  {Tamura}}]{oht16}
{Ohta}, Y., {et~al.} 2016, \pasj, 68, 53

\bibitem[{{Paardekooper} \& {Mellema}(2006)}]{paa06}
{Paardekooper}, S.-J., \& {Mellema}, G. 2006, \aap, 453, 1129

\bibitem[{{Pinilla} {et~al.}(2012){Pinilla}, {Birnstiel}, {Ricci}, {Dullemond},
  {Uribe}, {Testi}, \& {Natta}}]{pin12}
{Pinilla}, P., {Birnstiel}, T., {Ricci}, L., {Dullemond}, C.~P., {Uribe},
  A.~L., {Testi}, L., \& {Natta}, A. 2012, \aap, 538, A114

\bibitem[{{Price} \& {Bate}(2007)}]{pri07}
{Price}, D.~J., \& {Bate}, M.~R. 2007, \mnras, 377, 77

\bibitem[{{Reg{\'a}ly} {et~al.}(2012){Reg{\'a}ly}, {Juh{\'a}sz}, {S{\'a}ndor},
  \& {Dullemond}}]{reg12}
{Reg{\'a}ly}, Z., {Juh{\'a}sz}, A., {S{\'a}ndor}, Z., \& {Dullemond}, C.~P.
  2012, \mnras, 419, 1701

\bibitem[{Storn \& Price(1997)}]{sto97}
Storn, R., \& Price, K. 1997, Journal of Global Optimization, 11, 341

\bibitem[{{van der Marel} {et~al.}(2016){van der Marel}, {Cazzoletti},
  {Pinilla}, \& {Garufi}}]{van16}
{van der Marel}, N., {Cazzoletti}, P., {Pinilla}, P., \& {Garufi}, A. 2016,
  \apj, 832, 178

\bibitem[{{van der Marel} {et~al.}(2013){van der Marel}, {van Dishoeck},
  {Bruderer}, {Birnstiel}, {Pinilla}, {Dullemond}, {van Kempen}, {Schmalzl},
  {Brown}, {Herczeg}, {Mathews}, \& {Geers}}]{van13}
{van der Marel}, N., {et~al.} 2013, Science, 340, 1199

\bibitem[{{Weidenschilling}(1977)}]{wei77}
{Weidenschilling}, S.~J. 1977, \apss, 51, 153

\bibitem[{{Willson} {et~al.}(2017){Willson}, {Kraus}, \& {Kluska}}]{wil17}
{Willson}, M., {Kraus}, S., {Kluska}, J., {Monnier}, D., {Cure}, M., {Sitko}, M., {Aarnio}, A., {Ireland}, M., {Rizzuto}, A., {Hone}, E., {Kreplin}, A., {Andrews}, S., {Calvet}, N., {Espaillat}, C., {Fukagawa}, M., {Harries}, T.J., {Hinkley}, S., {Kanaan}, S., {Muto}, T., \& {Wilner}, D.J., \aap, 2017 (submitted)

\bibitem[{{Zhu} {et~al.}(2012){Zhu}, {Nelson}, {Dong}, {Espaillat}, \&
  {Hartmann}}]{zhu12}
{Zhu}, Z., {Nelson}, R.~P., {Dong}, R., {Espaillat}, C., \& {Hartmann}, L.
  2012, \apj, 755, 6

\end{thebibliography}

\end{document}